\documentclass[aps,prd,twocolumn,nofootinbib,floatfix]{revtex4-1}
\usepackage{setspace}
\usepackage{amsthm}
\theoremstyle{definition}

\usepackage{bbold}
\usepackage{graphicx}
\usepackage{dcolumn}
\usepackage{multirow}
\usepackage{subfigure}
\usepackage{times,mathptm}
\usepackage{float}
\usepackage{color}
\usepackage{amsmath,amsfonts}
\usepackage{mathptmx}
\usepackage{mathrsfs}
\usepackage{bbm}
\usepackage{bm}
\usepackage{xfrac}
\newcommand{\beq}{\begin{equation}}
\newcommand{\eeq}{\end{equation}} 
\newcommand{\bea}{\begin{eqnarray}}
\newcommand{\eea}{\end{eqnarray}} 
\newcommand{\Om}{\Omega}

\newcommand{\E}{\mathcal{E}}
\newcommand{\V}{{\cal V}}

\newcommand{\Sc}{S$_\text{c}$\ }

\renewcommand{\d}{\delta}

\newcommand{\ophi}{\overline{\phi}}

\newcommand{\p}{\psi}
\newcommand{\op}{\overline{\psi}}
\renewcommand{\b}{\beta}

\newcommand{\tr}{\text{Tr}}

\newcommand{\vx}{{\vec{x}}}
\newcommand{\vy}{{\vec{y}}}
\newcommand{\vz}{\vec{z}}

\newcommand{\m}{\mu}
\newcommand{\pbar}{\overline{\psi}}
\newcommand{\q}{\overline{q}}
\newcommand{\g}{\gamma}

\newcommand{\s}{\sigma}
\renewcommand{\k}{\kappa}

\renewcommand{\th}{\theta}

\newcommand{\dg}{\dagger}
\newcommand{\non}{\nonumber}

\newcommand{\rf}[1]{(\ref{#1})}
\newcommand{\ra}{\rightarrow}
\newcommand{\pa}{\partial}
\renewcommand{\vec}[1]{\bm #1}
\usepackage{ulem}

\begin{document}

\title{Separation-of-charge confinement and the Higgs transition in SU(3) gauge Higgs theory} 

\bigskip
\bigskip

\author{Jeff Greensite and Hou Y. Yau}
%\singlespacing
\affiliation{Physics and Astronomy Department \\ San Francisco State
University  \\ San Francisco, CA~94132, USA}
\bigskip
\date{\today}
\vspace{60pt}
\begin{abstract}

\singlespacing
 
With SU(3) gauge Higgs theory as an example, we examine critically the idea
that the confinement property in an SU(N) gauge-Higgs theory, with the Higgs field in
the fundamental representation, persists in an unbroken SU(N-1) subgroup.

 \end{abstract}

%
%\pacs{11.15.Ha, 12.38.Aw}
%
%\keywords{xxx}
%
\maketitle
 
\singlespacing

\section{\label{Intro} Introduction}

    It has been argued elsewhere \cite{Greensite:2020nhg} that the transition to the Higgs phase of an SU(N) gauge Higgs theory, with the Higgs field in the fundamental representation of the gauge group, is characterized by the spontaneous breaking of the global center subgroup of the gauge group, with an associated non-local order parameter which is closely analogous to the Edwards-Anderson order parameter for a spin glass \cite{Edward_Anderson}.  In the absence of a massless phase, the transition to the Higgs phase is accompanied by
a transition in confinement type, from ``separation-of-charge'' (S$_\text{c}$) confinement in the confinement phase, to a weaker  ``color'' (C) confinement property in the Higgs phase.  SU(3) gauge Higgs theory is a good testing ground for these assertions, in particular because of the idea that in this theory the SU(3) gauge symmetry is broken to SU(2), so that in some way the confinement property of the SU(3) theory is retained in a subgroup.  The purpose of this article is to examine this idea critically; the object is to study whether the concept
of separation-of-charge confinement applies to color charges (static ``quark'' sources) in color directions orthogonal to the color orientation of the Higgs field.

   In section \ref{theory} we review the order parameter for the confinement-to-Higgs transition presented in ref.\ \cite{Greensite:2020nhg}, define and contrast the \Sc and C varieties of confinement, and expand on the nature of the problem we address. %In that section we
%will also explain why reference to  unitary gauge  in discussions of the Brout-Englert-Higgs (BEH) mechanism
%can be a little misleading, at least in the lattice formulation.   
Section \ref{numbers} is devoted to the results of our
numerical simulations of SU(3) gauge Higgs theory concerning symmetry breaking and confinement.  Section \ref{conclude} contains our conclusions.

\section{\label{theory} Confinement in gauge Higgs theories}

   In this article the expression ``gauge Higgs theory'' will refer specifically to SU(N) gauge theories with a single Higgs field in the fundamental representation of the gauge group.

\subsection{Separation of charge confinement}

   The property of confinement in a gauge theory with either no matter fields, or with only matter fields
of zero N-ality, can be formulated in this way:  Let us consider, in lattice regularization, physical states of the form
\beq
    | \Psi_V  \rangle \equiv  \op^a(\vx) V^{ab}(\vx,\vy;U) \p^b(\vy) | \Psi_0 \rangle  \ ,
\label{PV}
\eeq
where $\Psi_0$ is the ground state and  $\op, \p$ are operators creating static fermion-antifermion sources, transforming in the
fundamental representation of the gauge group, at positions $\vx,\vy$.  Superscripts are color indices. The operator $V^{ab}(\vx,\vy;U)$ is
a bicovariant functional of the lattice gauge field $U$, transforming under a gauge transformation $g$ as follows:     
\beq
    V^{ab}(\vx,\vy;U) \ra   g^{ac}(\vx) V^{cd}(\vx,\vy;U) g^{\dg db}(\vy) \ .
\eeq 
If the energy expectation value above the vacuum energy
\beq
      E_V(R) =   \langle \Psi_V | H | \Psi_V \rangle - {\cal E}_{vac} 
\eeq
diverges to infinity as $R=|\vx-\vy| \ra \infty$, regardless of the choice of $V$, then the gauge theory
is said to be confining.  Of course there is one particular $V$, a flux tube state of some kind, which minimizes 
of the energy of a static quark-antiquark system, with $E_V(R) \approx \s R$ at large $R$.

    The proposal in \cite{Greensite:2017ajx} is that the very same condition can be used as a criterion for confinement in
a gauge theory with matter fields in the fundamental representation of the gauge group.  We call this
property ``separation of charge'' or ``S$_\text{c}$'' confinement.  We are considering physical states containing
isolated sources whose color charge is not screened by the matter field, and for this purpose it is understood
that the $V(\vx,\vy;U)$ operator is a functional of the gauge field alone, with no dependence on matter fields.  
In QCD, $q \q$ states of this
kind would contain, in addition to color electric fields collimated into flux tubes, also detectable abelian electric fields emanating from fractional electric charges at points 
$\vx,\vy$, rather than emanating from a set of integer charged
particles.  The point is that while physical states containing color charges unscreened by matter fields may
be challenging to produce in practice and, if produced, would decay very rapidly into ordinary hadrons,
states of this kind do exist in the physical Hilbert space, and it is reasonable to consider how their energy
varies as the separation between the color sources increases.  Indeed, if we imagine pair production followed by a very rapid separation
of the quark and antiquark, then we expect that momentarily the physical state of the separated $q \q$ pair
would have the form \rf{PV}.  In the \Sc phase the energy of the unscreened state increases with separation, and this is the mechanism which underlies the existence of linear Regge trajectories in QCD.  Of course such states exist only momentarily, until 
string breaking sets in.  In a phase without the \Sc property there is no reason to expect that the optimal $\Psi_V$ would be associated with linear Regge trajectories, so the loss of linear Regge trajectories is to be expected in a transition from a \Sc phase to a C onfinement phase.

%The only difference between
%the \Sc criterion in pure gauge theories, and the \Sc criterion in gauge theories with matter fields, is that in the latter case the state $\Psi_V$ of minimal energy is not the state of lowest energy in a $q\q$ system at large $R$. The lowest energy states are, of course, states where the color charge is screened by matter fields, and there is no long range color electric field emanating from the static sources.  These are not $\Psi_V$ states, and their energies tend to a finite constant a large quark separation.

    The alternative to \Sc confinement, for SU(N) gauge theories with matter in $D=2+1$ and $3+1$ dimensions,
is ``color'' or ``C'' confinement, meaning that the asymptotic spectrum consists of color neutral particles.
This is a much weaker condition than \Sc confinement, and in fact it holds true in gauge Higgs theory
in the Higgs regime, where there are no linear Regge trajectories or metastable flux tubes whatever.
There will still exist $V(\vx,\vy,U)$ operators such that $E_V(R)$ diverges with $R$.  An example is a Wilson line running between points $\vx,\vy$.  However, in the C confining phase (which we will identify with the
Higgs phase) there also exist  $V(\vx,\vy,U)$ operators such that $E_V(R)$ tends to a finite limit at large $R$.

    It can be shown that a transition from \Sc confinement (the ``confining'' phase) to C confinement (the
``Higgs'' phase) must exist in the SU(N) gauge Higgs phase diagram \cite{Greensite:2018mhh}.  It is natural to ask whether this
transition is accompanied by the breaking of some symmetry.

\subsection{Charged states and global center gauge symmety}
 
     A state which is charged with respect to some symmetry of the Hamiltonian is a state which transforms covariantly, as a non-singlet,
under those symmetry transformations.  So we might naively expect a state which is charged with respect to the gauge group
to transform under gauge transformations, e.g.\ a state like $\pbar^a(x) \Psi_0$, where $\Psi_0$ is the ground state.  But states of that
type violate the Gauss law constraint, and are hence unphysical.   Thus we are looking for a state of the form $\pbar^a(x)  \xi^a(x) \Psi_0$
which is invariant under infinitesimal gauge transformations, thereby satisfying the Gauss law, but still non-invariant under some subgroup
of the gauge group.

     Charged states of that type can be constructed; they are states which are charged with respect to the global center subgroup of
 the gauge group, i.e.\ which transform under space-invariant gauge transformations $g(x)=g$ where $g$ is a center element.  Note that transformations in this global subgroup will not transform the gauge field.  Our first example of a charged state is a static electric charge coupled to the quantized electromagnetic field.
 The lowest energy eigenstate of this system is as follows \cite{Dirac:1955uv}:
 \beq
          |\Psi_{\text{chrg}}\rangle  =  \op(\vx) \rho(\vx) |\Psi_0\rangle \ ,
\eeq
where
\bea
            \rho(\vx;A) &=& \exp\left[-i {e\over 4\pi} \int d^3z ~ A_i(\vz) {\pa \over \pa z_i}  {1\over |\vx-\vz|}  \right]  \ .
\eea
The operator $\rho(\vx;A)$ is an example of what we have called a ``pseudomatter'' field, namely, a non-local functional of the gauge field
which transforms like a matter field at point $\vx$, except under transformations in the global center subgroup (GCS) of the gauge group.
It is in fact obvious that such an operator is invariant under the GCS, because the gauge field itself is invariant under such transformations.  
In the abelian case, consider a U(1) gauge transformation $g(\vx) = e^{i\th(\vx)}$ on a  time slice, with $\th(x) = \th_0 + \tilde{\th}(x)$, where $\th_0$ is the zero mode of $\th(x)$.  The ground state $\Psi_0$ is invariant under this transformation, and $\pbar(x) \ra \pbar e^{-i\th(x)}$, but
\beq
      \rho(\vx;A) \ra e^{i\tilde{\th}(\vx)} \rho(\vx,A) \ ,
\eeq
As a result, $| \Psi_{\text{chrg}}\rangle$ is not entirely gauge invariant, but transforms as  
$|\Psi_{\text{chrg}}\rangle \ra e^{-i\th_0}  |\Psi_{\text{chrg}}\rangle$.
In other words, it is covariant under transformations in the U(1) global center subgroup of the U(1) gauge group.

   If the theory contains a single-charged scalar field, then we may construct neutral states, invariant under the GCS, such as
\beq
          |\Psi_{\text{neutral}}\rangle  =  \op(\vx) \phi(\vx) |\Psi_0\rangle \ .
\eeq
Providing the global center symmetry is unbroken, charged and neutral states are necessarily orthogonal in the confined and
massless phases.  But this is no longer true if the U(1) GCS is spontaneously broken, in which case we lose the sharp distinction
between charged and neutral states.

      All of this extends to non-abelian gauge theories.  In the case of SU(N) gauge Higgs theories, the GCS is the global
$Z_N$ center subgroup of the gauge group.\footnote{In the special case of SU(2) gauge Higgs theory, there is a larger global SU(2)
symmetry, known as ``custodial symmetry,'' which transforms the Higgs but not the gauge field.  This symmetry includes the GCS symmetry.}  Then we may construct charged states in lattice gauge theory of the form
\beq
  |\Psi_{\text{chrg}}\rangle  =  \op^a(x) \xi^a(x;U) |\Psi_0\rangle \ ,
 \eeq
 where $\xi^a(x;U)$ is a non-local functional of the link variables only which transforms like a field in the fundamental representation
 of the gauge group {\it except} under transformations in the GCS, i.e.\ it is a pseudomatter field.   Thus, under
 a gauge transformation $g(x) = z \mathbb{1}, ~ z \in Z_N$,
 \beq 
       |\Psi_{\text{chrg}}\rangle \ra z^*   |\Psi_{\text{chrg}}\rangle \ .
 \eeq
 Examples of non-abelian pseumatter fields include the eigenstates $\zeta^a_n(\vx;U)$ of the covariant Laplacian operator, $-D^2 \zeta_n = \lambda_n \zeta_n$, where\footnote{Pseudomatter fields of this type can be combined to construct transformations to gauges  which avoid the Gribov ambiguity, cf.\ \cite{Vink:1992ys}.}
\bea
   (-D^2)^{ab}_{xy} &=&  \sum_{k=1}^3 \bigg[2 \delta^{ab} \delta_{xy} 
   - U_k^{ab}(x) \delta_{y,x+\hat{k}}    - U_k^{\dg ab}(x-\hat{k}) \delta_{y,x-\hat{k}}  \bigg]  \ . \non \\
\label{laplace}
\eea

     As in the abelian theory, one can also construct neutral states in which the
charge of the fermion is entirely shielded by the Higgs field:
$|\Psi_{\text{neutral}}\rangle  =  \op^a(\vx) \phi^a(\vx) |\Psi_0\rangle$, and these are invariant with respect to the GCS.  Assuming
that the GCS is not spontaneously broken, then it is obvious that $\langle |\Psi_{\text{neutral}}|\Psi_{\text{chrg}}\rangle = 0$.
The phase in which this is no longer true, and there is no longer a sharp distinction between charged and neutral states, is the Higgs
phase.

\subsection{Order parameter}

It is not hard to construct an order parameter for the spontaneous breaking of a GCS.   Consider
\bea
  e^{-H(U,\phi)} &\equiv& \Psi_0^2[U,\phi] \non \\
  Z[U] &=& \int D\phi(x) ~  e^{-H(U,\phi)} 
\eea
where it is understood that all fields are defined on some time slice.  Introduce
\beq
   \ophi(x;U) = {1\over Z(U)} \int D\phi ~ \phi(x)  e^{-H(U,\phi)} 
\eeq
Then the GCS is spontaneously broken in the background $U$ field if $ \ophi(x;U) \ne 0$.  Since the background $U$ breaks translation
symmetry we can expect that the spatial average of $\ophi(x;U)$ will vanish even in the broken phase.  So it is convenient to
introduce
\beq
          \Phi[U] = {1\over V} \sum_x | \ophi(x;U)|
\label{max}
\eeq
Now we take the expectation value $\Om = \langle \Phi[U] \rangle$.  If $\Om \ne 0$, then the global center subgroup of the gauge
group is spontaneously broken in the full theory.  So $\Om$ is the desired order parameter.\footnote{There are strong similarities  
between $\Om$ and the Edwards-Anderson order parameter \cite{Edward_Anderson} for a spin glass.
See \cite{Greensite:2020nhg}.}

     The central result of \cite{Greensite:2020nhg} is that the spontaneous
breaking of the global center subgroup of the gauge group, as detected by the order parameter $\Om$,  is accompanied by a transition from \Sc to C confinement.\footnote{Here we have ignored some technicalities.  Formally there is no spontaneous symmetry breaking of a global symmetry in a finite volume; the rigorous procedure is to introduce a term, proportional to some parameter $h$,  which explicitly breaks the global symmetry.  We then
evaluate $\Omega$ first taking the thermodynamic limit, and then the $h\ra0$ limit.  The rigorous statement is that the GCS is broken
if  $\Omega$ is non-zero after the two limits, taken in this order.  We refer the reader to \cite{Greensite:2020nhg} for a detailed discussion of the breaking term, but for the numerical treatment these formalities will not be necessary.}  Briefly, when $\ophi(x;U)$ is nonzero, it may be
used to define a gauge in which $\ophi(x;U)$ points everywhere in a given color direction.  Then one can construct a $V(\vx,\vy;U)$ operator from the product of gauge transformations to this gauge, one transformation at point $x$, and the conjugate at point $y$.  It can be shown that for this $V$ operator, $\Psi_V$ is no longer orthogonal to neutral states in which color is screened by the Higgs field, and  $E_V(R)$ tends to a finite constant as $R\ra \infty$.  Then by definition the broken symmetry phase is not an \Sc confining phase.  For details, cf.\ \cite{Greensite:2020nhg}.

\subsection{GCS and \Sc confinement in SU(3)}

\vspace{-5pt}
This finally brings us to the question we would like to address. 
In an SU(N) gauge Higgs theory,
it is stated in the standard texts, e.g.\ \cite{Srednicki:2007qs}, that the Higgs mechanism ``breaks'' the SU(N) gauge symmetry
to SU(N-1).  More precisely, what can be seen perturbatively in unitary gauge is that the Lagrangian
in this gauge supplies a mass term for some of the gauge bosons, leaving the bosons corresponding
to the generators of an SU(N-1) group massless, at least at the perturbative level.  The implication is that, in $D\le 4$ dimensions, confinement doesn't disappear entirely; it should remain for the quark components which transform among themselves
via the SU(N-1) group.  But what is really meant by the word ``confinement'' in this situation?  We have already defined confinement in a gauge theory with matter fields
as \Sc confinement, as opposed to the weaker property of C confinement.  Then the question is whether \Sc confinement can be seen in some way in the SU(N-1) subgroup.  For example, in an SU(3) gauge theory with a unimodular constraint $\phi^\dg \phi = 1$,
in a unitary gauge which sets
\beq
\phi(x) = \left[ \begin{array}{c} 0 \cr 0 \cr 1 \end{array} \right] \ ,
\eeq
 then the fermion components orthogonal to the $\phi$ color direction are simply the first two color components
 of the $\p^a(x)$ field, namely $a=1$ and $a=2$.  These are the color components which, in unitary gauge,
 are said to be ``confined'' by the unbroken SU(2) gauge symmetry.  We will refer to these components as 
 ``quarks'' $q$.  The gauge invariant generalization is the fermion field multiplied by a color projection operator $P^{ab}(x)$
 \bea
 q^a(x) &=& [\d^{ab} - \phi^a(x)\phi^{\dg b}(x)] \p^b(x) \non \\
            &=&  P^{ab}(x) \p^b(x) \ ,
 \label{qdef}
 \eea
 and we consider the energy $E^q_V(R)$ of physical states of the form
 \beq 
    | \Psi'_V  \rangle \equiv  \q^a(\vx) V^{ab}(\vx,\vy;U) q^b(\vy) | \Psi_0 \rangle  \ .
\eeq
If in the Higgs phase $E^q_V(R)$ diverges with $R$ regardless of $V$, then 
there is \Sc  confinement of quarks transforming in the  ``unbroken'' sector.  On the contrary, if we can find
some $V$ operators such that $E^q_V(R)$ converges to a finite constant at large $R$, then it is hard to make sense
of the claim that the quarks are confined, in any sense other than the property of C confinement which holds for all fermion components, and not just the quarks.
 
\section{\label{numbers}Numerical results}

    The SU(3) lattice action is
\beq
     S = - {\b\over 3} \sum_{\text{plaq}} \text{Re}\tr[UUU^\dg U^\dg ] - \g \sum_{x,\m} \text{Re}[\phi^\dg(x) U_\m(x)\phi(x+\hat{\m}] \ ,
\label{S}
\eeq
where we impose, for convenience, the unimodular condition $|\phi(x)|=1$.  At $\g=\infty$ and unitary gauge this reduces to SU(2) gauge theory, since all but the SU(2) degrees of freedom are frozen.  In this limit, the theory is certainly confining.  But are the quark degrees of freedom also confined at finite $\g$ and, if not, how is confinement regained as $\g\ra \infty$?

We begin with simulations at $\b=6.0$ and a variety of $\g$ values on $16^4$ lattice volumes, followed by simulations
at $\b=3.6$, which is in the strong-coupling regime of pure SU(3) lattice gauge theory.  A few details about the link updates,
which greatly improve efficiency at large $\g$, are provided in the Appendix.

\subsection{Symmetry breaking transition}

 The first
question is at which value of $\g$ the symmetry breaking transition takes place, according to the order parameter $\Om$, and this is determined numerically as follows:
The SU(3) gauge and scalar fields are updated in the usual way, but each data-taking sweep (separated by
100 update sweeps) actually consists of a set of $n_{sym}$ sweeps in which the spacelike links $U_i(\vx,0)$ are held fixed on the $t=0$ time slice.  So data taking is, in a sense, a ``Monte Carlo in a Monte Carlo.''  Let $\phi(\vx,t=0,n)$ be the scalar field at site $\vx$ on the $t=0$ time slice at the N-th sweep.  Then we compute $\overline{\phi}(\vx,U)$ from the average over $n_{sym}$ sweeps 
\beq
             \overline{\phi}(\vx,U) = {1\over n_{sym}} \sum_{n=1}^{n_{sym}} \phi(\vx,0,n) \ ,
\eeq
and  $\Phi_{n_{sym}}(U)$ from \rf{max}.  Here it is important to indicate the dependence
on $n_{sym}$. Then the procedure is repeated, updating links and the scalar field together, followed by another computation of $\Phi_{n_{sym}}(U)$ from a simulation with spatial links at $t=0$ held fixed, and so on.     Averaging the 
$\Phi_{n_{sym}}(U)$ obtained by these means results in an estimate for $\langle \Phi_{n_{sym}} \rangle$.  
Since $\Phi_{n_{sym}}(U)$ is a sum
of moduli, it cannot be zero.  Instead, on general statistical grounds, we expect
%~\footnote{One must keep in mind that for the symmetry breaking parameter $h=0$ and finite volume used in the numerical computation, $ \langle \Phi \rangle$ would actually vanish at ${n_{sym} \ra \infty}$, since a symmetry cannot break in a finite volume.  The proper order of limits is first $V \ra \infty$, then $n_{sym} \ra \infty$.  Nevertheless, for $n_{sym}$ not too large,  \rf{fit} is a good fit to the data, and the extrapolation should be reliable.} 
\beq
           \langle \Phi_{n_{sym}} \rangle =  \Omega  + {\k \over \sqrt{n_{sym}}} \ ,
\label{fit}
\eeq
where $\k$ is some constant.  By computing  $\langle \Phi_{n_{sym}} \rangle$ in independent runs at a range of $n_{sym}$ values, and fitting the results to \rf{fit}, we obtain an estimate for $\Om$ at any point in the  $\b,\g$ plane of lattice couplings.

\begin{figure}[htb]
\includegraphics[scale=0.6]{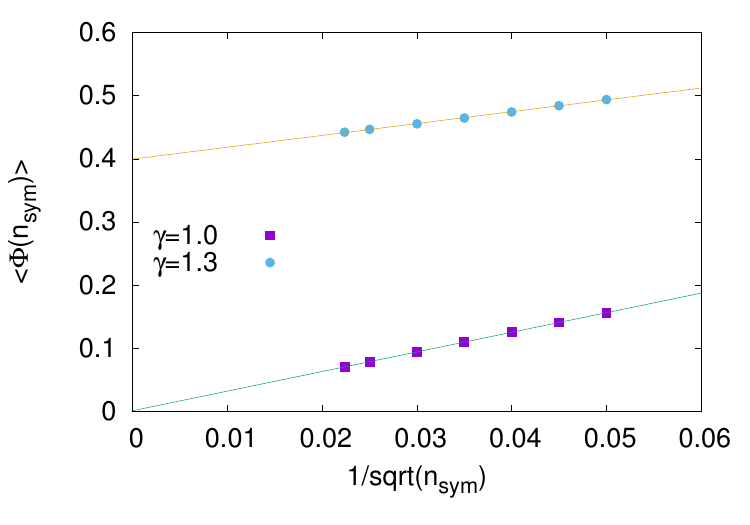}
 \caption{Computation of the order parameter $\langle \Phi(n_{sym})\rangle$ vs.\ $1/\sqrt{n_{sym}}$ where $n_{sym}$ is number of Monte Carlo sweeps (see text).  Also shown is the extrapolation to $n_{sym}=\infty$.  Data was taken at  $\b=6$ for $\g=1.0$ and $\g=1.3$.  
 Extrapolation to $\langle \Phi \rangle =0$ indicates that the system is in the \Sc confined phase, while extrapolation to 
 $\langle \Phi \rangle > 0$ means that the system is in the Higgs phase.}
  \label{nsym}
\end{figure}

\begin{figure}[htb]
 \includegraphics[scale=0.6]{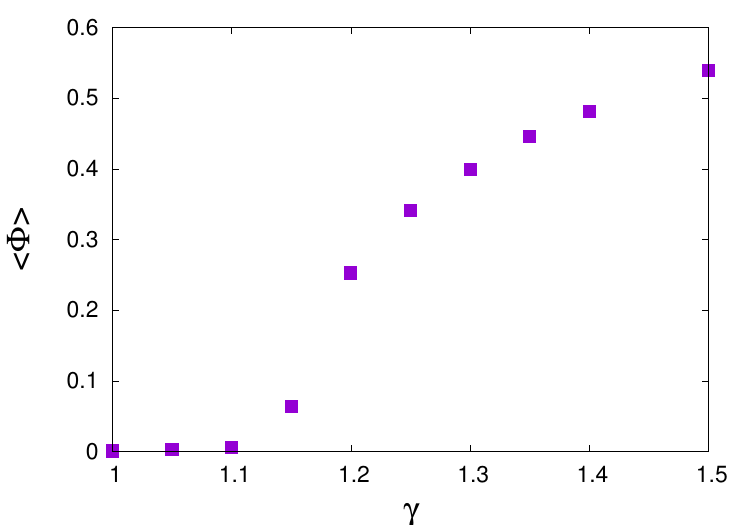}
 \caption{Order parameter $\langle \Phi \rangle = \Om$ at $n_{sym}=\infty$, computed by extrapolations of the kind shown in Fig.\ \ref{nsym}, vs.\ $\g$ at
 fixed $\b=6.0$.}
 \label{Phi}
\end{figure}

   In principle, in a finite volume, one should include an explicit GCS symmetry breaking term in the action, proportional to some parameter $h$, and then compute $\Om$ by first taking the thermodynamic and then the $h\ra 0$ limit.  For the numerical simulations this procedure
is not really necessary, and we may set the symmetry breaking parameter $h$ to zero, providing
$n_{sym}$ is not too large. Of course, at $h=0$ and finite volume, $\langle \Phi \rangle$ must vanish
at $n_{sym} \ra \infty$, since a symmetry cannot break at finite volume.  Nevertheless, for $n_{sym}$ in the range
we have used, \rf{fit} turns out to be a good fit to the data, and the extrapolation to $n_{sym}=\infty$ should be reliable.  Figure \ref{nsym} shows typical data for $\langle \Phi(n_{sym})\rangle$ vs.\
$1/\sqrt{n_{sym}}$ at $\g=1.0$ and $\g=1.3$, with $\langle\Phi\rangle$ determined from the intercept of the straight line
fit with the y-axis.  The transition point, where $\langle\Phi\rangle$ moves away from zero, appears to be for $\g$ somewhere
between $\g=1.1$ and $\g=1.15$, as shown in Fig.\ \ref{Phi}.

\begin{figure}[htb]
 \includegraphics[scale=0.6]{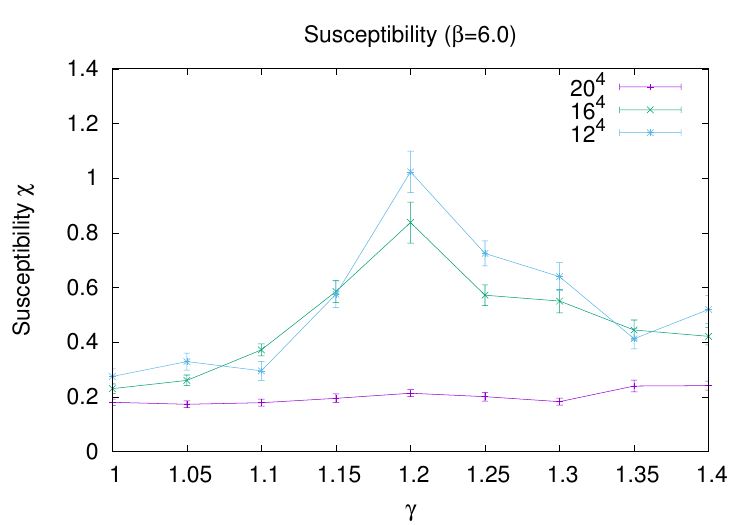}
 \caption{Gauge invariant link susceptibilities $\chi$ vs.\ $\g$ at $\b=6$ and several lattice volumes.}
 \label{suscept}
\end{figure}

   It is worth noting that there does not appear to be any thermodynamic transition at $\b=6.0$ at any $\g$, as we
see from a plot of the link susceptibility $\chi_L$ vs.\ $\g$ in Fig.\ \ref{suscept},  where
\beq
         L = \sum_{x,\m} \text{Re}[\phi^\dg(x) U_\mu(x) \phi(x)]
\eeq
and
\beq
        \chi_L = 4 \V  (\langle L^2\rangle - \langle L \rangle^2)
\eeq
where $\V$ is the lattice volume. The work of 
\cite{Osterwalder:1977pc,Fradkin:1978dv,Banks:1979fi} tells us that the confinement and Higgs regions are not entirely isolated from one another by a line of thermodynamic transition, so while there could have been such a thermodynamic transition at $\b=6$ and some $\g$, this is not required.  Note that 
$\Phi(U)$ is a non-local functional of the gauge field, and the expectation value of non-local functionals can have non-analytic behavior even when the free energy is analytic in that region.

\begin{figure*}[t!]
\subfigure[~]  % caption for subfigure a
{\label{Ag1p15}
 \includegraphics[scale=0.6]{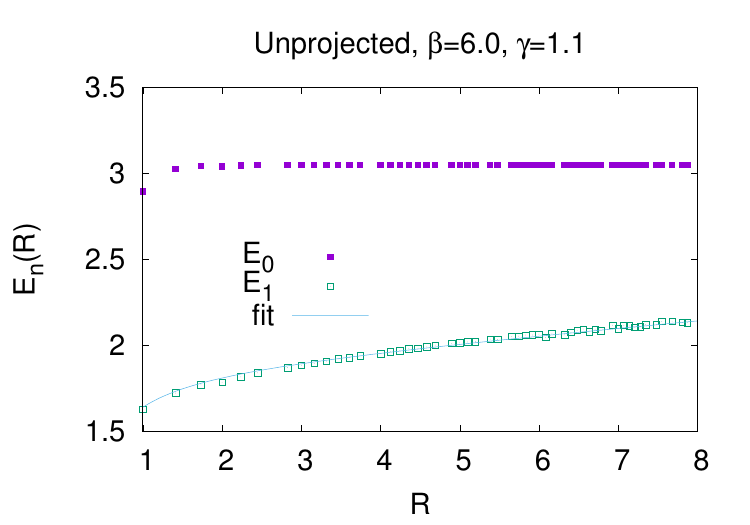}
 } 
\subfigure[~]
{\label{Ag1p2}
 \includegraphics[scale=0.6]{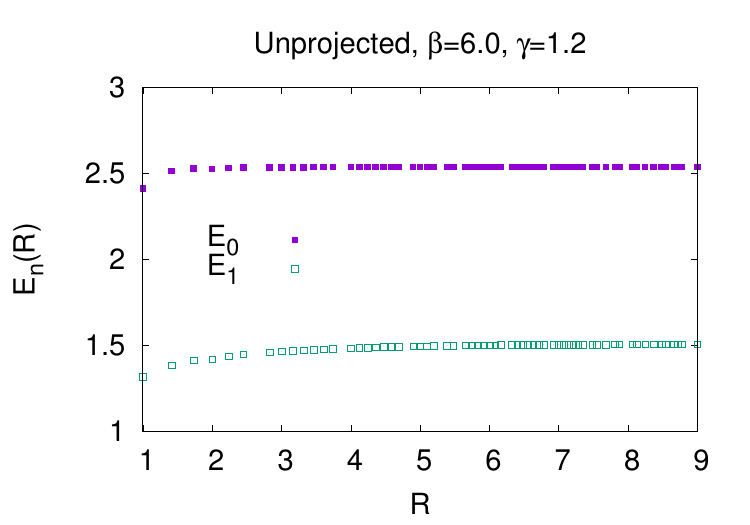}
}
 \caption{Energy expectation values $E_0(R)$ and $E_1(R)$ for unprojected fermion states just below (subfigure (a)) and just above (subfigure (b)) the confinement to Higgs transition.  $E_0$ corresponds to a state where the fermionic color charges are neutralized by the Higgs field, while all $E_n(R)$ with
 $n>0$ are derived from operators $V(\vx,\vy;U)$ built from eigenstates of the lattice Laplacian operator.  The fact that $E_1(R)$ diverges with $R$
 is required in the confinement phase, while the fact that $E_1(R)$ converges to a constant, while not required, implies that the system
 is in the Higgs phase.  The term "unprojected" means that the fermions are not projected to the quark states, with color orthogonal
 to the Higgs field. }
 \label{A}
 \end{figure*}

\begin{figure*}[htb]
\subfigure[~]  % caption for subfigure a
{\label{Eg1p1}
 \includegraphics[scale=0.45]{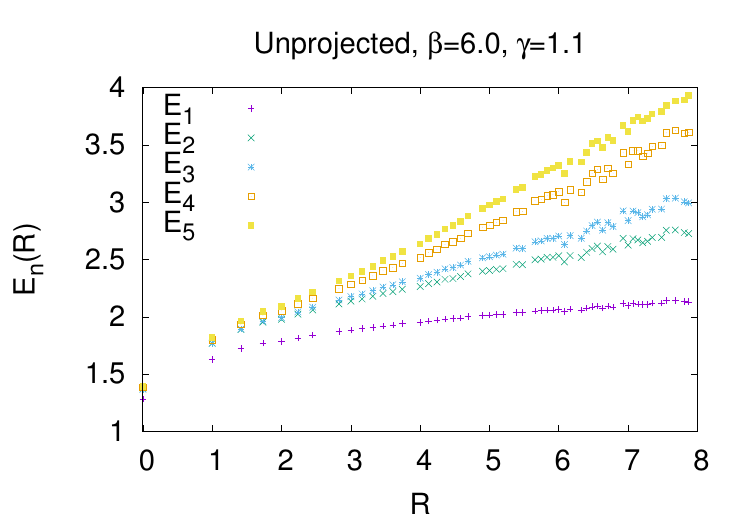}
 } 
\subfigure[~]
{\label{Eg1p2}
 \includegraphics[scale=0.45]{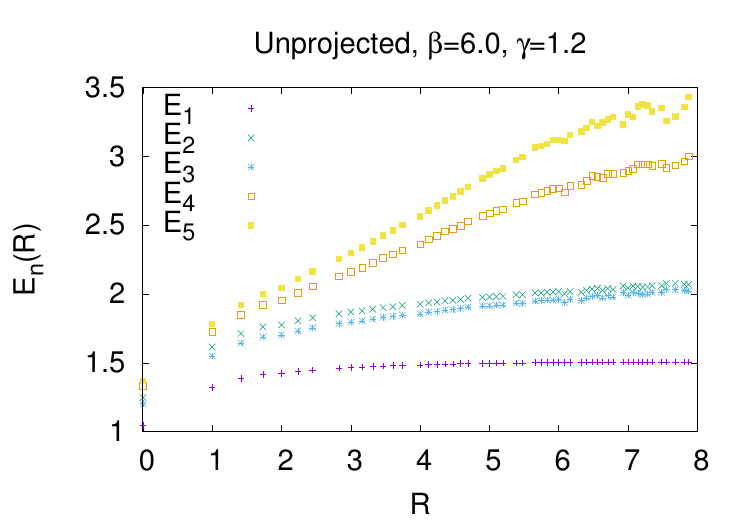}
}
\subfigure[~]
{\label{Eg1p8}
 \includegraphics[scale=0.45]{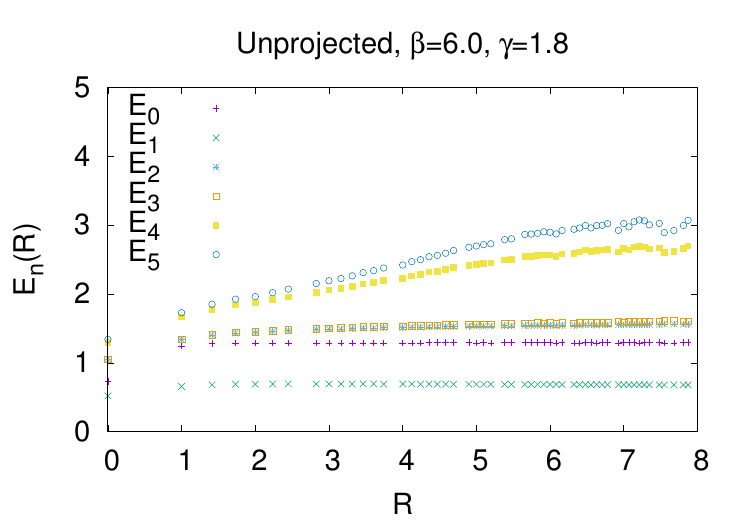}
 }
 \caption{Same as Fig.\ \ref{A} for $E_n$ with $n=1-5$ at (a) $\g=1.1$, in the confined phase; and (b) $\g=1.2$ , (c) $\g=1.8$ both in
 the Higgs phase.  Note that while only $E_1$ goes ``flat'' at $\g=1.2$, also $E_2$ and $E_3$ are flat at the higher $\g=1.8$ coupling, where
 $E_0$ is also displayed for comparison.}
 \label{E}
 \end{figure*}
\subsection{$\mathbf E_V$ across the transition}

We first consider the \Sc criterion with no projection to the quark fields.  Let
\beq
         | \Psi_n(R)\rangle = Q_n(R) |\Psi_0\rangle \ ,
\eeq
where
\bea
          Q_0(R) &=& [\op^a(\vx) \phi^a(\vx)] ~\times ~  [\phi^{\dg b}(\vy) \p^b(\vy)]   \non \\
         Q_n(R) &=& [\op^a(\vx) \zeta_n^a(\vx)] ~\times~  [\zeta_n^{\dg b}(\vy) \p^b(\vy)] ~~~~(n>0)  \ .
\label{Qn}
\eea
For $n>0$, the $Q_n$ are of the form $\op V \p$ with factorizable $V(\vx,\vy;U)$ operators 
\beq
V^{ab}(\vx,\vy;U) = \zeta_n^a(\vx)\zeta_n^{\dg b}(\vy) \ , 
\eeq
where $\zeta_n$ is the n-th eigenstate of the covariant Laplacian operator
\rf{laplace}. The fermions are static, and propagate only in the time direction. For continuous time, the energy 
expectation value above the vacuum energy $\E_0$ is
\bea
     E_n &=& -\lim_{t\ra 0} {d\over dt} \log\left[\langle \Psi_n| e^{-(H-\E_0)t} |\Psi_n\rangle \right] \non \\
            &=& - \lim_{t\ra 0} {d\over dt}\log\left[ \langle Q_n^\dg(R,t) Q_n(R,0) \rangle \right]  \ , 
\eea
where $Q_n(R,t)$ is shown in \rf{Qn}, with operators on the right hand side defined on time slice $t$.
The corresponding expression in discretized time is
\beq
         E_n = - \log \left[ { \langle Q_n^\dg(R,1) Q_n(R,0) \rangle \over \langle Q_n^\dg(R,0) Q_n(R,0)\rangle} \right] \ .
\eeq
After integrating out the heavy quark fields, and dropping an $R$-independent constant, we have, for $n>0$,
 \bea
  & & \lefteqn{\langle Q_n^\dg(R,1) Q_n(R,0) \rangle} \non \\
   & & \qquad = \bigg\langle \left[  \zeta_n^{\dg}(\vx,0)  
                               U_0(\vx,0) \zeta_n(\vx,1) \right]  [ \zeta_n^{\dg}(\vy,1)U^{\dg}_0(\vy,0) 
                               \zeta_n(\vy,0) ] \bigg\rangle \non \\
 & & \lefteqn{\langle Q_n^\dg(R,0) Q_n(R,0) \rangle} \non \\                               
   & & \qquad = \bigg\langle \left[  \zeta_n^{\dg}(\vx,0)  
                              \zeta_n(\vx,0) \right] \left[   \zeta_n^{\dg}(\vy,0) 
                               \zeta_n(\vy,0) \right] \bigg\rangle  \ .
\label{QQ1}         
 \eea
 The expressions are the same for $n=0$, with $\zeta_n$ replaced by the Higgs field $\phi$.
 
    In Fig.\ \ref{A} we see $E_0(R)$ and $E_1(R)$ just below ($\g=1.1$) and just above $(\g=1.2)$ the
symmetry breaking transition.  $E_0(R)$ is the energy of a pair of separated color neutral objects, and of
course we do not expect any significant dependence on $R$, as we see in the figure.  In the symmetric
phase, the claim is that any $E_V(R)$ diverges with $R$, and this is certainly true for $E_1(R)$, as seen
in Fig.\ \ref{Ag1p15}. The data for $E_1(R)$ is fit to the form
\beq
    E^{fit}(R) = a + b R - {\pi\over 12R} \ ,
\label{Efit}
\eeq
and the coefficient of the linearly rising term is $b \approx  0.039$. This can be compared to the asymptotic string tension of the pure ($\g=0$) SU(3) gauge theory at $\b=6.0$, which is $\s=0.048$ \cite{Cardoso:2011hh}.  It should be understood that at any finite $\g$ the asymptotic string tension, as extracted from, e.g., Wilson loops or Polyakov line correlators, is zero, due to string breaking.  This was the motivation to construct the \Sc criterion.  In the
broken phase there is no prediction; $E_V(R)$ might diverge, or it might converge to a constant,  depending on $V$.
The claim is only that there must exist {\it some} $V$ such that $E_V(R)$ becomes flat at large $R$.  In fact $E_1(R)$
has this convergence property only a little past the transition, as seen in  Fig.\ \ref{Ag1p2}.  If we were ignorant
of the order parameter $\langle \Phi \rangle$, the behavior of $E_1(R)$ at $\g \ge 1.2$ would be sufficient to establish that the system  is in the 
Higgs phase in that region.

    It is instructive to look at $E_n(V)$ for other $n$, below and above the symmetry breaking transition,
and the results are seem in Fig.\  \ref{E}.  As predicted, all $E_{n>0}(R)$ rise with $R$ in the symmetric region (Fig.\ 
\ref{Eg1p1} at $\g=1.1$. In the Higgs region at $\g=1.2$, $E_1(R)$ flattens out, while the other $E_{n>0}$ continue
to rise linearly (Fig.\ \ref{Eg1p2}).  At still higher $\g$, we find that more of the $E_n(R)$ become flat (Fig.\  \ref{Eg1p8}).

 \subsection{Quark $\mathbf E_V$ across the transition}
 
 \begin{figure*}[h!]
\subfigure[~]  % caption for subfigure a
{\label{Pg1p1}
 \includegraphics[scale=0.6]{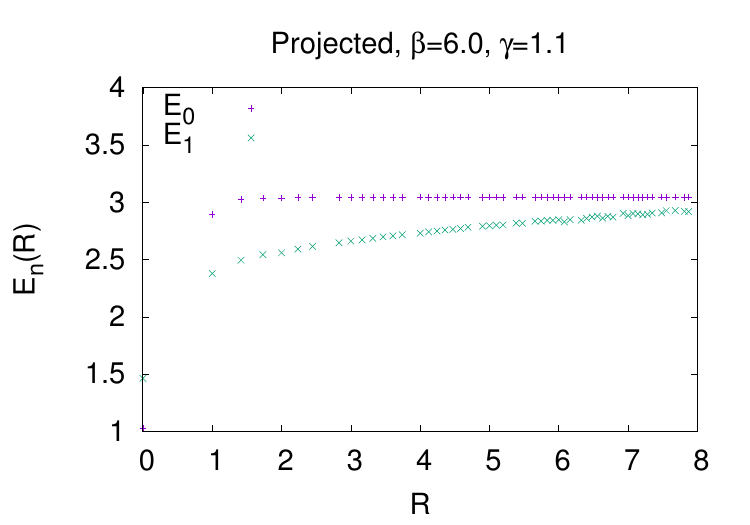}
 } 
\subfigure[~]
{\label{Pg1p2}
 \includegraphics[scale=0.6]{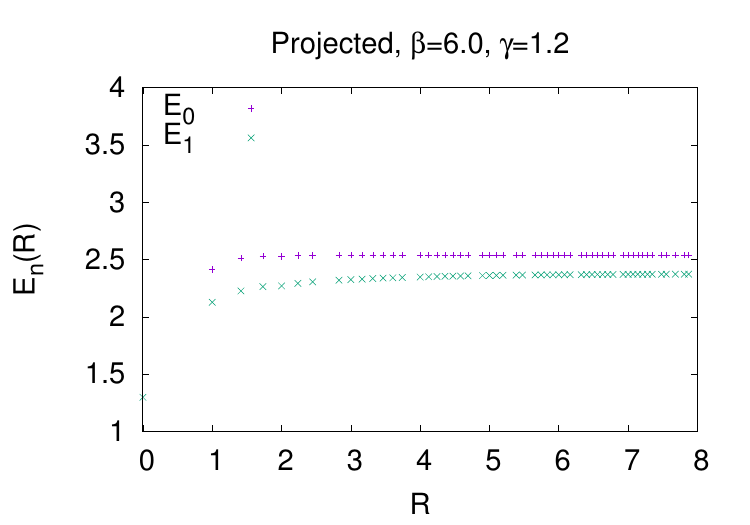}
}
 \caption{Same as Fig.\ \ref{A}, this time for fermions projected to quark components orthogonal to the color direction
 of the Higgs field.  The important point here is that in contrast to subfigure (a) in the confined phase, $E_1(R)$ goes flat
 in the Higgs phase (subfigure (b)), meaning that the quarks which transform into themselves under an SU(2) subgroup do not have
 the \Sc confinement property.}
 \label{A1}
 \end{figure*}
 
\begin{figure*}[h!]
\subfigure[~]  % caption for subfigure a
{\label{Pg1p1all}
 \includegraphics[scale=0.45]{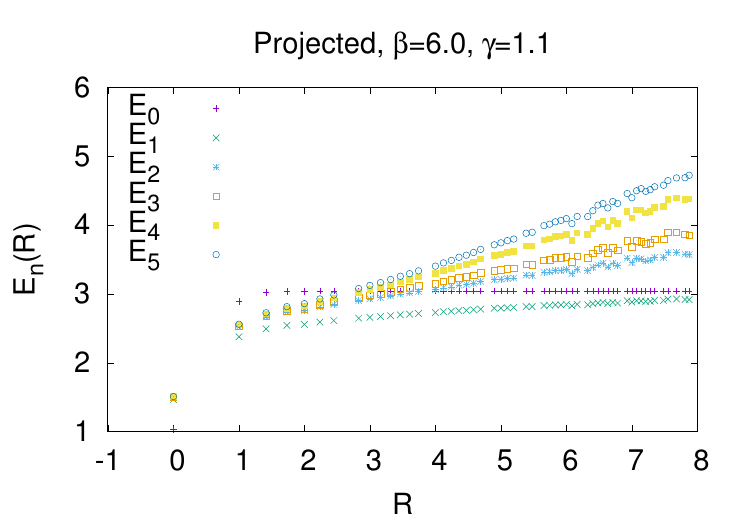}
 } 
\subfigure[~]
{\label{Pg1p2all}
 \includegraphics[scale=0.45]{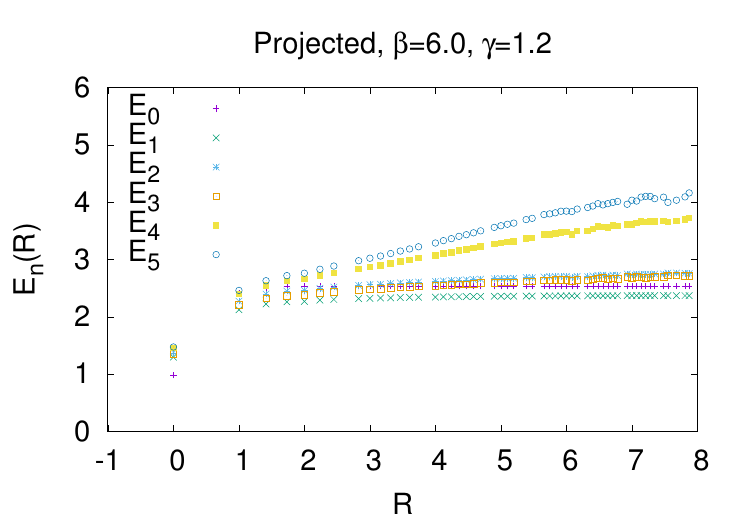}
}
\subfigure[~]
{\label{Pg1p8all}
 \includegraphics[scale=0.45]{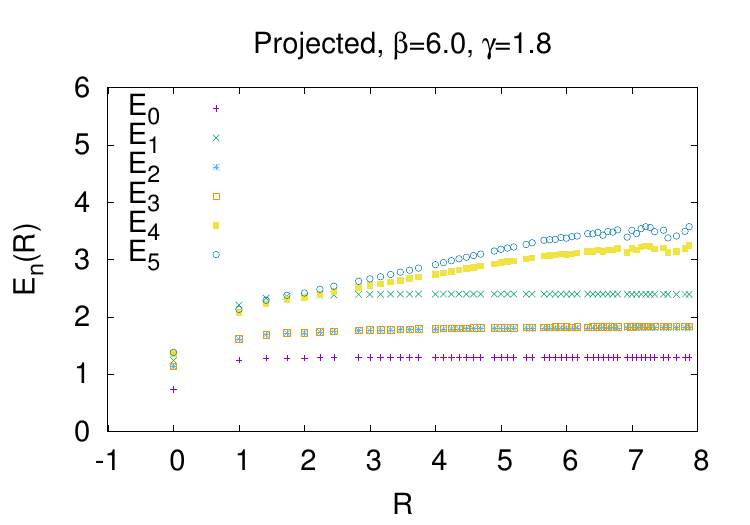}
 }
 \caption{Same as Fig.\ \ref{E} but for fermions projected to quarks as described in the text.}
 \label{E1}
 \end{figure*}
 
     The intent of using $\op V \p$ operators to create physical states is to exclude ``broken string'' states, in which the color of the fermion is neutralized by the scalar.  In the confinement phase the corresponding $E_V(R)$ diverges with $R$, and a few examples were presented in the previous section.  The question of interest is whether the
Higgs phase could still be a confinement phase for the quark operators $q$, whose color orientation is, by definition, orthogonal to that of the scalar field.   Therefore we consider just replacing the $\op \p$ operators 
by $\q q$ operators, and computing the energy expectation
values $E^q_n(R)$ of states
\bea
         | \Psi^q_n(R)\rangle &=& \q(\vx) V_n(\vx,\vy) q(\vy) |\Psi_0\rangle \non \\
                                       &=&  Q^q_n(R) |\Psi_0\rangle \ ,
\eea
where this time
\beq
Q^q_n(R) = [\q^a(\vx) \zeta_n^a(\vx)] ~\times~  [\zeta_n^{\dg b}(\vy) q^b(\vy)] 
\eeq
and $Q^q_0 = 0$ by definition.  The superscript $q$ indicates that there is a color projection to the quark operators.Then we have
\beq
         E^q_n(R) = - \log \left[ { \langle {Q^q}_n^\dg(R,1) Q^q_n(R,0) \rangle \over \langle {Q^q}_n^\dg(R,0) Q^q_n(R,0)\rangle} \right]           
\eeq
as before, but this time, after integrating out the heavy quark fields,
 \bea
 & & \lefteqn{\langle {Q^q}_n^\dg(R,1) Q^q_n(R,0) \rangle} \non \\ 
 & & ~ = \bigg\langle \left[  \zeta_n^{\dg a}(\vx,0) P^{ab}(\vx,0) 
                               U^{bc}_0(\vx,0) P^{cd}(\vx,1) \zeta_n^d(\vx,1) \right] \non \\
 & & ~ \times \left[   \zeta_n^{\dg e}(\vy,1) P^{ef}(\vy,1) U^{\dg fg}_0(\vy,0) P^{gh}(\vy,0) 
 \zeta_n^{h}(\vy,0) \right] \bigg\rangle \ ,
\label{QQ1'}         
 \eea
and
 \bea      
    \langle {Q^q}_n^\dg(R,0) Q^q_n(R,0) \rangle &=& \bigg\langle \left[  \zeta_n^{\dg a}(\vx,0) P^{ab}(\vx,0) 
                      \zeta_n^b(\vx,0) \right] \non \\
                      &\times& \left[   \zeta_n^{\dg e}(\vy,0) P^{ef}(\vy,0) \zeta_n^{f}(\vy,0) \right] \bigg\rangle  \ . \non \\
\label{QQ0}
\eea
where $P^{ab}$ is the color projection operator defined in \rf{qdef}.
Then the question is whether the energies $E^q_n(R)$ of the $\q q$ states satisfy the \Sc criterion in the Higgs phase.
If so, this would supply a gauge-invariant meaning to the claim that confinement somehow exists, in the Higgs phase,
in an ``unbroken'' SU(2) subgroup.

At $\b=6.0$ there is no evidence for this claim.  Figure \ref{A1} shows the results of a calculation of $E_0(R)$ (as defined
above) along with $E^q_1(R)$ using quark sources, both in the confinement ($\g=1.1$) and Higgs ($\g=1.2$) phases.  As in Fig.\ \ref{A}, $E^q_1(R)$ rises linearly in the confined phase, with a string tension (extracted from a fit to
\rf{Efit}) of $\s\approx 0.04$.   But in the Higgs phase $E^q_1(R)$ is flat, with no discernable linear component, as found in Fig.\ \ref{A}
for the unprojected fermion field.  This implies the absence of \Sc \ confinement for quarks in the Higgs phase.
For completeness we show, in Fig.\ \ref{E1}, the results for $E^q_n(R)$ in a range of $n\ge 1$, at $\g=1.1,1.2,1.8$.

   However, the interpretation of these results is not entirely clear, due to the fact that a pure SU(2) gauge theory on
a $16^4$ lattice, at $\b=6.0$, would be deep in the deconfined phase of the theory.  Suppose we go to unitary gauge
and assume that, in the Higgs phase, the gauge bosons of the SU(2) subgroup are approximately decoupled from the other gauge bosons of the SU(3) theory.  Denote by $\tilde{U}_\m$ the $2\times 2$ sub-matrix of components $U^{ab}_\m$ with
indices $a,b$ in the range $1,2$.  Then the part of the Lagrangian involving only the degrees of freedom of the 
SU(2) subgroup is
\beq
       S_{eff} \approx - {\b \over 3} \sum_{plaq} \text{Re}\tr[UUU^\dg U^\dg] 
                    = - {\b_{eff} \over 2} \sum_{plaq} \tr[\tilde{U} \tilde{U} \tilde{U}^\dg \tilde{U}^\dg]  \ ,
\eeq
with an effective SU(2) coupling
\beq
\b_{eff}={2\over 3}\b
\label{beff}
\eeq
which, at $\b=6$, corresponds to $\b_{eff} = 4$.  But at this effective
SU(2) coupling, on a $16^4$ lattice, the pure SU(2) theory is in the deconfined phase.
Hence the lack of \Sc \ confinement might be attributable to finite size effects.  To eliminate the source of
ambiguity, we repeat our calculation for $\b=3.6$, which is in the strong coupling regime of pure SU(3) gauge theory,
but which corresponds to $\b_{eff}=2.4$.  On a $16^4$ volume, at this effective coupling, pure SU(2) gauge theory 
is in the confined phase, and the test of \Sc \ confinement is less susceptible to finite size effects.

\subsection{Numerical results at $\bf \b=3.6$}

\begin{figure}[h!]
 \includegraphics[scale=0.6]{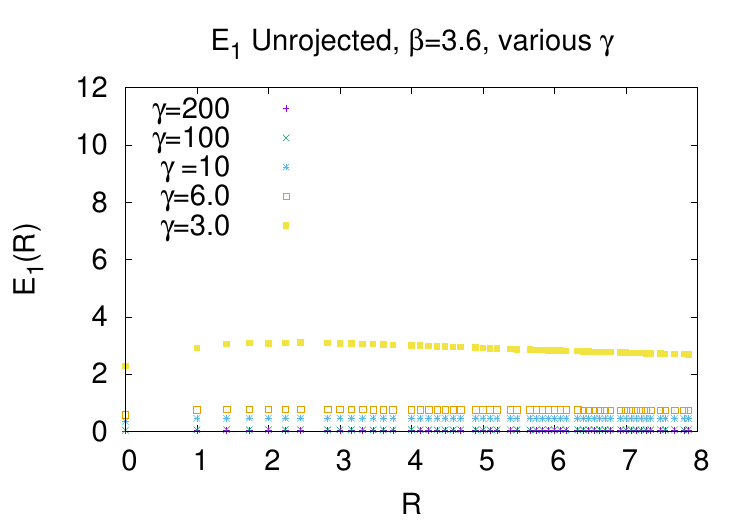}
 \caption{$E_1(R)$ vs.\ $R$ for unprojected fermions in the Higgs phase, at  $\b=3.6$ and various $\g$ ranging up to $\g=200$.  Note that
 $E_1(R)$ converges to zero with increasing $\g$.}
 \label{E1uproj}
\end{figure}

\begin{figure}[h!]
 \includegraphics[scale=0.6]{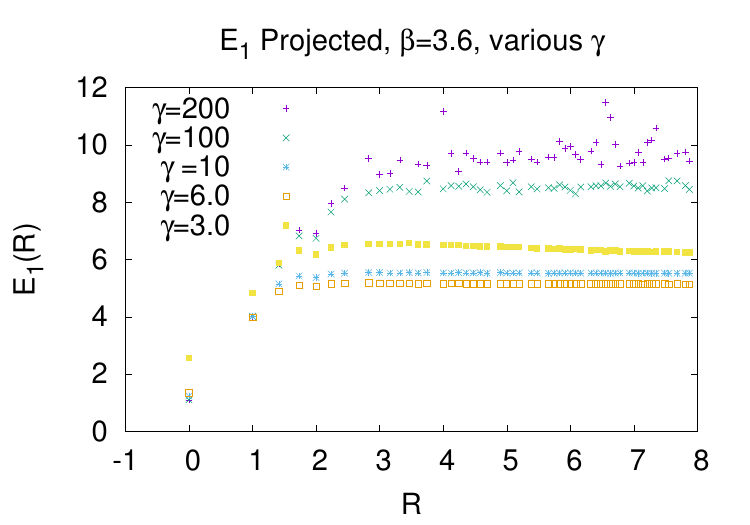}
 \caption{$E_1(R)$ vs.\ $R$ for quarks in the Higgs phase, at  $\b=3.6$ and various $\g$ ranging up to $\g=200$.  At each $\g$
 there is no \Sc confinement.  However, the behavior with increasing $R$ is opposite to the unprojected case, in that the value at which $E_1(R)$ flattens out tends to increase with $\g$. }
 \label{E1proj}
\end{figure}
 
\begin{figure}[h!]
 \includegraphics[scale=0.6]{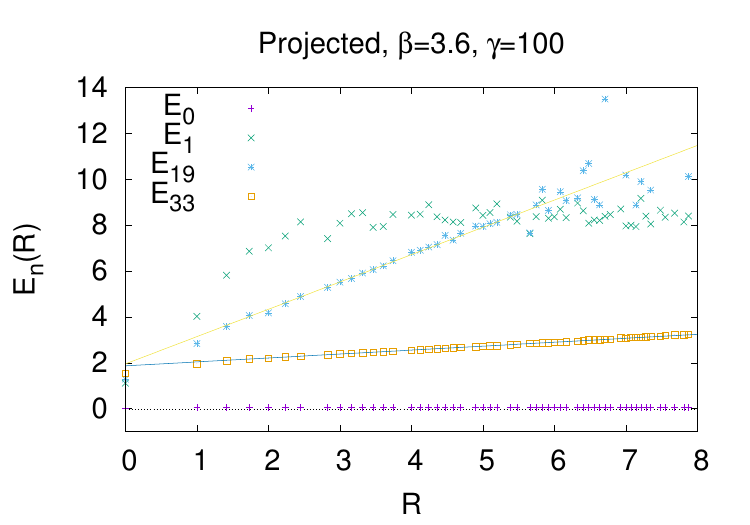}
  \caption{$E^q_n$ at selected values of $n$ at $\b=3.6$ and a large $\g=100$.   We observe that while $E_1(R)$ rises abruptly and flattens out at $E_1 = 8$, $E_{19}(R)$ rises linearly in the range shown, with string tension $\s_{19}=1.19(4)$, while $E_{33}(R)$ also
rises linearly in this range, with string tension $\s_{33}=0.172(1)$.  For comparison, in the $\g\ra\infty$ limit, the coupling of the effective SU(2) pure gauge  theory is $\b_{SU()2)}=2.4$, and the corresponding string tension of the minimal energy flux tube state is $\s=0.071$ \cite{Heller} in lattice units.}
 \label{Pb36g100}
\end{figure}

   At $\g=\infty$ we expect that the SU(3) gauge Higgs theory reduces exactly to a pure
SU(2) gauge theory, with an effective SU(2) Wilson coupling $\b_{eff}$ of \rf{beff}, and this theory must
have the property of \Sc \ confinement for the quark degrees of freedom.
The question is whether this property exists at finite $\g$ in the Higgs phase and, if not, how the
\Sc \ property is approached in the continuum limit.   For this purpose it is sufficient to compute the $E_n(R)$
at $\b=3.6, ~\b_{eff} = 2.4$ and at $\g$ values which are deep within the Higgs regime.

\begin{figure}[h!]
\includegraphics[scale=0.6]{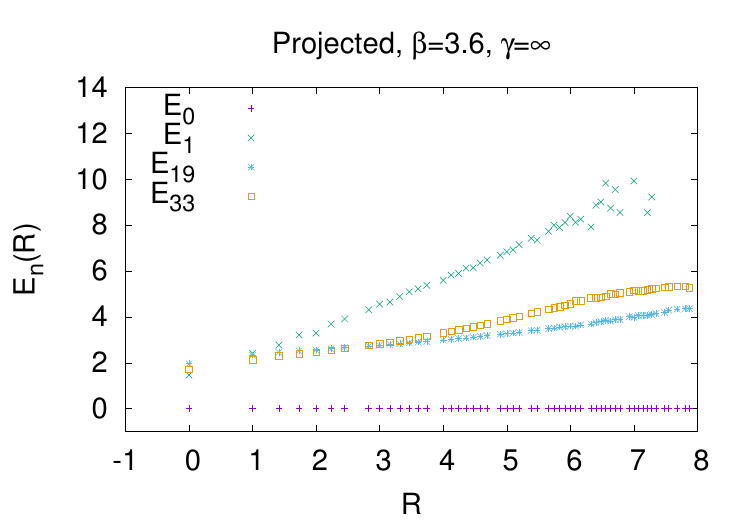}
\caption{$E^q_n$ at selected values of $n$ at $\b=3.6$ at $\g=\infty$.} 
\label{B}
\end{figure}

   The data shown in Fig.\  \ref{E1uproj} for $E_1(R)$ at various $\g$, without projection to the quark states, shows that there is
there is no evidence for \Sc \ confinement, even at very large $\g$ values.  In fact, these energies are not only $R$ independent,
but drop to zero as $\gamma$ increases.  This is unsurprising, since there is no projection here to the ``unbroken'' SU(2) subgroup.

   The quark projection data is more interesting.  From the data it is fairly clear that, as at $\b=6.0$,  there is also no \Sc confinement of quarks in the Higgs phase at $\b=3.6$, as we see from plots of $E_1(R)$ in Fig.\ \ref{E1proj}.  At any of the $\g$ values shown in the figure, up to a very large value of $\g=200$, the \Sc quark confinement criterion is violated for $E^q_V(R)$ by the $V$ operator built from the
lowest Laplacian eigenmode
\beq
         V(\vx,\vy,U) = \zeta_1(\vx) \zeta_1^\dg(\vy) \ .
\eeq
But this raises the question of how the $\g \ra \infty$ limit recovers \Sc \ confinement.  The first observation is that while
the energy $E^q_1(R)$ is seen to plateau at fairly small separations $R$, the value of $E^q_1(R)$ at the plateau
appears to increase with $\g$.  It is reasonable to conclude that the plateau rises to infinity at $\g \ra \infty$.  But this
is not sufficient; one would certainly expect there to be $V$ operators, e.g.\ flux tube states of some kind, such that
$E^q_V(R)$ rises linearly, at some slope compatible with the SU(2) string tension of the effective SU(2) theory, as $\g\ra \infty$.
And we indeed see evidence of this happening in Fig.\ \ref{Pb36g100}, where some of the
$E^q_n(R)$ at larger $n$ values appear to have exactly this behavior.   Of course these are not minimal energy flux tube states, so the
string tensions of the $E_n(R)$ must be greater than the string tension of the minimal energy flux tube state, which is 
$\s=0.071$ in the $\g=\infty$ limit at $\b_{eff}=2.4$.  

   It is interesting to compare Fig.\  \ref{Pb36g100} with the same results at $\g=\infty$.  In unitary gauge, this simply amounts to
initializing link variables to a $3\times 3$ unit matrix, and then only updating the upper left $2\times 2$ matrix (the first step of
link updates described in the appendix).  The results at $\b=3.6 \Longrightarrow \b_{eff}=2.4$ are shown in Fig.\ \ref{B}.  The
most striking difference is in the behavior of $E_1(R)$ in the two figures.   The distinction between the $\g=100$ and $\g=\infty$
data is not so dramatic for $n=19, 33$, but still noticeable.  The comparison suggests that although the $\g \ra \infty$ limit 
recovers confinement for the quarks, as expected, this limit is still not quite the same as $\g=\infty$.

\section{\label{conclude} Conclusions}

   ``Confinement'' is a word that must be carefully defined in an SU(N) gauge theory with matter in the fundamental representation
of the gauge group.  Here and elsewhere  \cite{Greensite:2020nhg,Greensite:2017ajx} we have argued that it is important to distinguish between color (C) confinement, which
means that there is a color neutral particle spectrum, and the much stronger condition of separation-of-charge (\Sc) confinement.  The latter condition means that the energy of physical states with separated color charges unscreened by matter fields increases
without limit as the color charge separation increases.  The transition from the confining to the Higgs phase of an SU(N) gauge theory corresponds to a transition from \Sc to C confinement.

   The question we have addressed in this article is whether, in an SU(N) gauge Higgs theory, the separation of charge property persists
in an SU(N-1) subgroup or, more precisely, for ``quark'' sources with color orthogonal to that of the scalar field, transforming among themselves, in a unitary gauge, via an SU(N-1) subgroup.  The answer which we find numerically for SU(3) gauge Higgs theory
is that \Sc confinement is lost in the Higgs phase also for the quarks transforming among themselves in the SU(2) subgroup.  Since it is certain that \Sc confinement must reappear in
a certain limit ($\g\ra \infty$ for the action in \rf{S}), it is of interest to see how \Sc confinement is regained in that limit.  To investigate
this, we have constructed quark-antiquark states with unscreened color, made gauge invariant by the use of eigenmodes of the
covariant Laplacian operator.  In the Higgs phase there are always quark-antiquark states which violate the \Sc condition, i.e.\ the
energy of such states rises to a value which is constant with charge separation.  This constant value, however, rises with $\g$.   There
are other gauge invariant states whose energy increases with charge separation, but which, for some range of quark separation,
is less than that constant value.  Thus the evidence suggests that, while there is strictly speaking no \Sc confinement at all in the Higgs phase, including for quark-antiquark states, the energy of states which violate the \Sc condition goes to infinity as $\g \ra \infty$,
leaving only states which satisfy the \Sc condition.

 \acknowledgments{This research is supported by the U.S.\ Department of Energy under Grant No.\ DE-SC0013682.}   

\bibliography{sym3}

 \appendix*
 \section{}
 
This is a brief note about a slightly modified procedure for link updates which are useful at large $\g$. 
We first fix to unitary gauge
\beq
          \phi(x) = \left[ \begin{array}{c}
                            0 \cr 0 \cr 1 \end{array} \right]
\eeq
and update link variables as follows:  From three stochastically generated SU(2) matrices with components $r_{ij},s_{ij}, t_{ij} $ we construct three SU(3) matrices $R,S,T$ where
\bea
          R &=&  \left[ \begin{array}{ccc}
                    r_{11} & r_{12} & 0 \cr
                    r_{21} & r_{22} & 0 \cr
                       0      &   0   &    1 \end{array} \right] \non \\
          S &=&  \left[ \begin{array}{ccc}
                    s_{11} &    0   & s_{12}  \cr
                      0       &    1   &  0        \cr
                    s_{21}  &   0   &   s_{22}  \end{array} \right] \non \\
          T &=&  \left[ \begin{array}{ccc}
                       1  &     0    & 0 \cr
                       0  &  t_{11} & t_{12} \cr
                       0  &  t_{21}   &  t_{22} \end{array} \right]                     
\eea
and from those obtain an updated link in two steps.  The average deviation of $R,S,T$ from the unit matrix is controlled
by a parameter $z$.  In the first step, we generate a trial link $=RU$ using a parameter $z=z_1$ for the $R$ matrix, and accept or reject according to the usual Metropolis algorithm.  In the second step the trial link is $RST U^{(1)}$, where  $U^{(1)}$ is the link obtained at the
first step, and this time the deviation of the $R,S,T$ matrices from the unit matrix is controlled by a parameter $z=z_2$.  Parameters
$z_1,z_2$ are chosen to give a 50\% acceptance at each of the two steps.  The reason for the two parameters is that in unitary gauge,
at large $\g$, the matrix components  $U^{ij}$ with $i,j=1,2$ may deviate significantly from a $2\times 2$ unit matrix,
while $U^{33} \approx 1$, and the remaining components are close to zero.  The first step only affects the $U^{ij}$ with $i,j=1,2$ components,
and therefore $R$ in the first step is allowed to deviate significantly from a unit matrix.  In the second step, which affects the remaining
components, $RST$ must be close to the unit matrix in order to have a reasonable acceptance rate.
   
\end{document}